\documentclass[runningheads]{llncs}
\usepackage[T1]{fontenc}
\usepackage{graphicx}
\usepackage{amsmath}
\usepackage{tabularx,booktabs}
\usepackage{subfigure}
\usepackage{hyperref}
\usepackage{float}

\begin{document}
\title{Generalist Segmentation Algorithm for Photoreceptors Analysis in \\ Adaptive Optics Imaging}
%
%
\author{Mikhail Kulyabin\inst{1} \and
Aline Sindel\inst{1} \and
Hilde R. Pedersen\inst{2} \and \\
Stuart Gilson\inst{2} \and
Rigmor Baraas\inst{2} \and
Andreas Maier\inst{1}}
\authorrunning{M. Kulyabin et al.}
%
\institute{Pattern Recognition Lab, Department of Computer Science, \\ Friedrich-Alexander-Universität Erlangen-Nürnberg, Erlangen, Germany \\ \email{mikhail.kulyabin@fau.de}\\ \and
National Centre for Optics, Vision and Eye Care, Faculty of Health and Social Sciences, University of South-Eastern Norway, Kongsberg, Norway}

\titlerunning{Generalist Segmentation Algorithm for Photoreceptors Analysis}

\maketitle              
\begin{abstract}
Analyzing the cone photoreceptor pattern in images obtained from the living human retina using quantitative methods can be crucial for the early detection and management of various eye conditions. Confocal adaptive optics scanning light ophthalmoscope (AOSLO) imaging enables visualization of the cones from reflections of waveguiding cone photoreceptors. While there have been significant improvements in automated algorithms for segmenting cones in confocal AOSLO images, the process of labeling data remains labor-intensive and manual. This paper introduces a method based on deep learning (DL) for detecting and segmenting cones in AOSLO images. The models were trained on a semi-automatically labeled dataset of 20 AOSLO batches of images of 18 participants for 0$^{\circ}$, 1$^{\circ}$, and 2$^{\circ}$ from the foveal center. F1 scores were 0.968, 0.958, and 0.954 for 0$^{\circ}$, 1$^{\circ}$, and 2$^{\circ}$, respectively, which is better than previously reported DL approaches. Our method minimizes the need for labeled data by only necessitating a fraction of labeled cones, which is especially beneficial in the field of ophthalmology, where labeled data can often be limited.
\end{abstract}
\section{Introduction}
Adaptive optics scanning light ophthalmoscopy (AOSLO) \cite{roorda2002adaptive} offers a noninvasive approach to achieve high-resolution, in vivo imaging of the cone photoreceptors (cones) mosaic in both healthy and diseased retinas \cite{wynne2021promises}. The AOSLO technique integrates an adaptive optics (AO) system within a scanning light ophthalmoscope (SLO) \cite{scoles2014vivo}. The AO system employs a wavefront sensor and an actuated mirror to measure and dynamically compensate for wavefront aberrations caused by the eye's inhomogeneous medium. While AO can be utilized with any ophthalmic imaging device requiring light passage into or out of the eye, it is predominantly used with SLOs due to its superior contrast and resolution capabilities. Multimodal AOSLO imaging captures three channels simultaneously (confocal, split-detection, and dark-field), each emphasizing different retinal structures. The confocal modality of AOSLO facilitates relatively explicit imaging of cones and rods \cite{scoles2014vivo}, presenting clinicians and researchers with quantifiable but complex retinal structural information \cite{pedersen2024multimodal}. Using this technology, one can obtain various quantitative measures of the cone mosaic from AOSLO images, such as cone density, spacing, and pattern regularity \cite{litts2017photoreceptor}, \cite{cooper2016evaluating}. Such quantities are useful for developing sensitive biomarkers for early diagnosis and monitoring of ocular and systemic disease progression.

Considering just the cones, peak foveal density can approach 200,000 cones per $mm^{2}$ \cite{curcio1990human}, making manual labeling impractical. On the other hand, existing automatic labeling techniques may not consistently enable the automatic identification of every cone within an image, particularly in the presence of blood vessels or when the image clarity is compromised. Furthermore, the challenge intensifies when examining retinal locations that are more eccentric from the fovea. 

Using the Voronoi algorithm, we cover the area from center-to-center of a cone detected in the confocal image \cite{litts2017photoreceptor}. As we move out from the foveal center, we move from an area with only cones and where the Voronoi cell is equal to the cone's size to areas with rods in between cones. This has already happened about 0.5$^{\circ}$$^{\circ}$from the fovea center. Thus, in areas with rods and cones, the Voronoi represents distances between cones but not their size. Classical methods, such as presented in work by Li and Roorda \cite{li2007automated}, which are currently used in contemporary works, rely on the optical fiber properties of cone photoreceptors. In practice, the algorithm can mislabel rods as cones. Therefore, it needs to be revised by a human expert. New algorithms should take this into account. Several algorithms have been previously developed to detect inner segments in split-detection images \cite{cunefare2018deep,sredar2021comparison}. In general, AOSLO split-detection images are semi-automatically analyzed to extract the location of cone photoreceptor cells within the images, with compulsory refinement by a medical expert. Creating a fully automatic method for the segmentation and detection of cones will significantly increase the possibilities of retinal research and reduce the workload of retinal researchers. This paper introduces a deep learning (DL) --based method for automatically detecting and segmenting the cones.

\section{Related Works}
Cellpose \cite{stringer2021cellpose} is a versatile, generalist algorithm for cell segmentation in microscopy images, regardless of the imaging modality or the type of cells being analyzed. It employs a DL model to identify cell boundaries, enabling automated and accurate segmentation of individual cells or nuclei across various applications. The algorithm uses a novel approach based on the concept of ``flows'' to capture cells' complex shapes and sizes, making it highly effective in different biological contexts. The term ``flows'' refers to the vector field that is generated for each pixel in the image, pointing towards the center of the cell to which that pixel belongs. Cellpose 2.0 \cite{pachitariu2022cellpose} is an updated version with a manual correction step for training custom models. However, it requires considerable effort to manually correct the detected polygons of multiple cells, which is significant for the number of receptor cells in AOSLO images (up to 200,000 cells per \(mm^{2}\)). Another segmentation method, PolarMask, is a single-shot, anchor-free convolutional neural network (CNN) framework designed for instance segmentation \cite{xie2020polarmask}. Unlike traditional instance segmentation methods that rely on bounding boxes or complex, multi-stage processes, PolarMask simplifies this by utilizing a polar representation to capture the shape of each object. It generates a center point for the object and then defines the segmentation boundary through a set of rays emanating from the center to the boundary in polar coordinates. This approach allows PolarMask to perform instance segmentation efficiently and accurately without the need for anchor boxes, reducing the complexity and computational demands of the task. StarDist is a novel image segmentation method optimized for microscopy images, particularly those of nuclei and cells, leveraging a shape-based approach to outline individual objects' boundaries \cite{schmidt2018}, \cite{weigert2020}. The core innovation of StarDist lies in its use of star-convex shapes for segmentation, where it predicts the distances from the center of an object to its boundary in a fixed set of directions, effectively capturing the often complex and irregular shapes of biological cells. This method is implemented through a DL framework, allowing it to learn from annotated training data and generalize well to new, unseen images. StarDist stands out for its ability to handle overlapping structures and varying shapes, making it highly effective for tasks where segmenting closely packed or irregularly shaped cells is critical. Its performance and efficiency make it a valuable tool for biomedical image analysis, facilitating advanced quantitative studies of cellular structures.

These methods are versatile and efficient computational tools for segmentation, demonstrating significant performance in various biological imaging contexts \cite{waisman2021automatic}, \cite{stevens2022stardist}; however, they are designed to generalize across different types of cells and imaging modalities by leveraging a unique representation of cell shapes. Despite its robustness and adaptability, applying them to segment structures derived from Voronoi diagrams may require modifications. Cunefare et al. \cite{cunefare2018deep} applied CCN to confocal AOSLO images to detect the cones, extracting the training patches using the Voronoi algorithm. However, the method does not involve the segmentation of the cells. The AOSLO images belonged to patients with achromatopsia disease — which have many inactive cones — and are, in fact, black regions on the images and are very different from the active cones in our dataset.

\section{Methods}

\subsection{Dataset}
\label{dataset}
\begin{figure*}[h]
  \centering
  \subfigure[\empty]{\includegraphics[height=6cm, width=6cm]{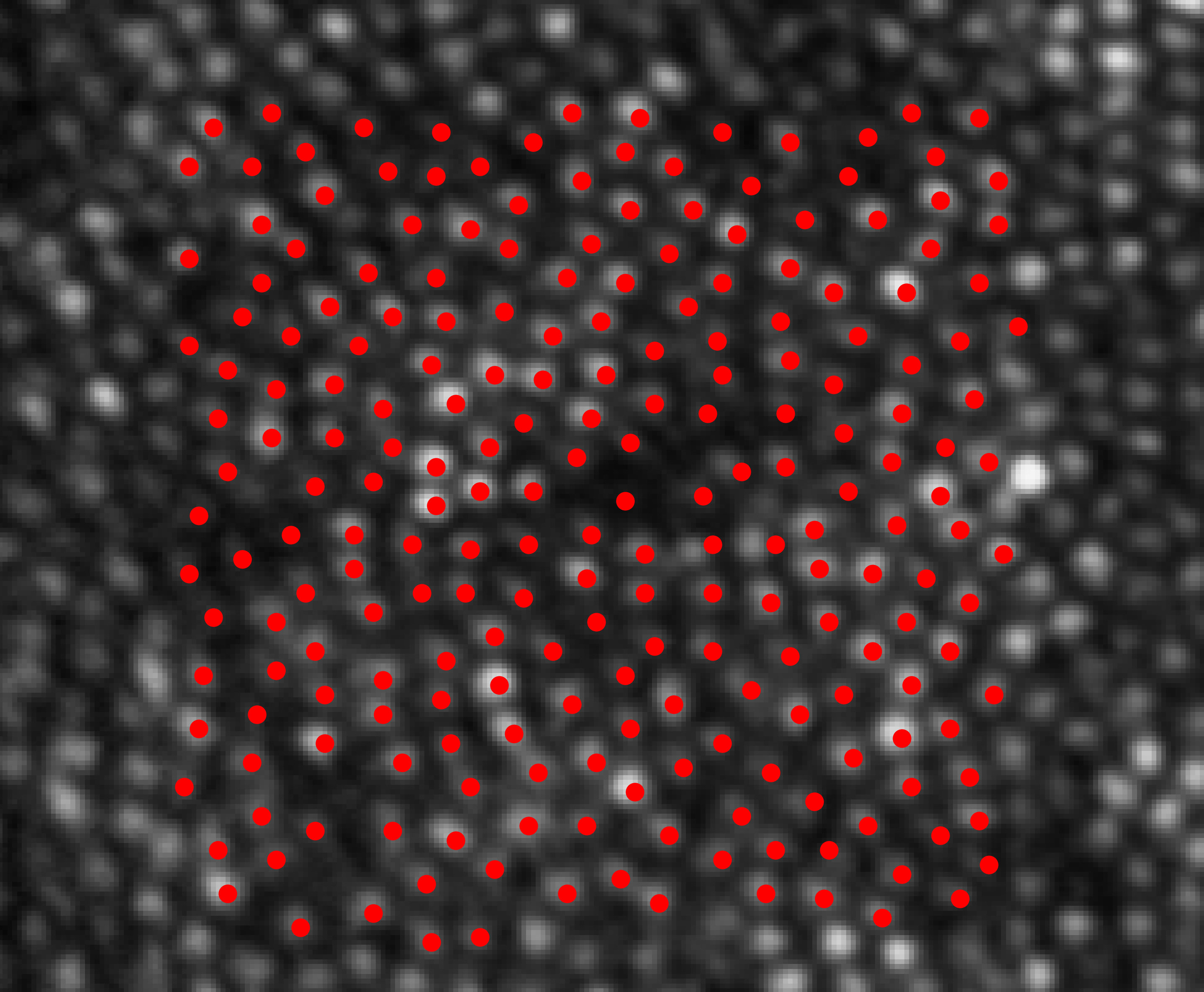}}
  \subfigure[\empty]{\includegraphics[height=6cm, width=6cm]{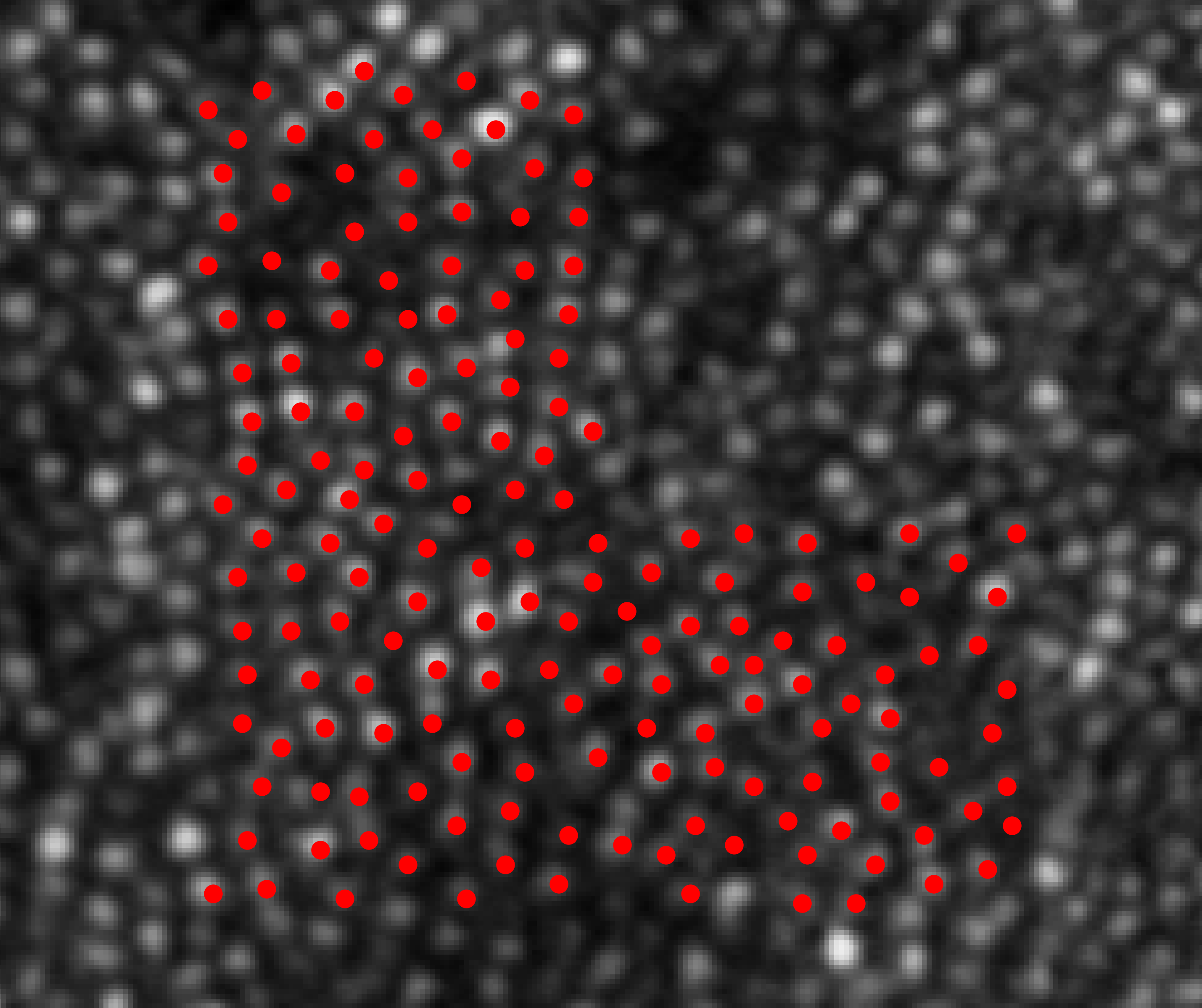}}
\caption{Two examples of confocal AOSLO images with labeled cone centers using the existing semi-automatic segmentation method \cite{li2007automated} followed by refinement by a medical expert.}
\label{fig:data}
\end{figure*}

In this study, we employed a semi-automatically labeled dataset of 20 AOSLO batches of images of 18 healthy participants with normal vision from a wide age range, 14–65 years, representing a wide scope of healthy retinas. Each batch consists of approximately 40 confocal images. The cone centers were first automatically identified with the classical method \cite{li2007automated}. Then, missed cones were manually added by a human expert or removed if they were mislabeled by the automatic algorithm, representing approximately 5\% of all cones across the dataset images. The data was split on the participant's level so that images belonging to one participant appeared only in one subset, with a train:test split ratio of 70:30. Therefore, we had 14 batches (540 images) of AOSLO images for training and 6 batches (240 images) for testing. Each image was cropped to $550\times550$ pixel resolution with 350 labeled cones on each image on average, for a total of 190k segmented cells in the training subset at the starting point. Fig.~\ref{fig:data} shows examples of confocal images with labeled cone centers.

\subsection{Human-in-the-loop approach}
\begin{figure}[h]
\begin{minipage}[b]{1.0\linewidth}
  \centering
  \centerline{\includegraphics[width=12cm]{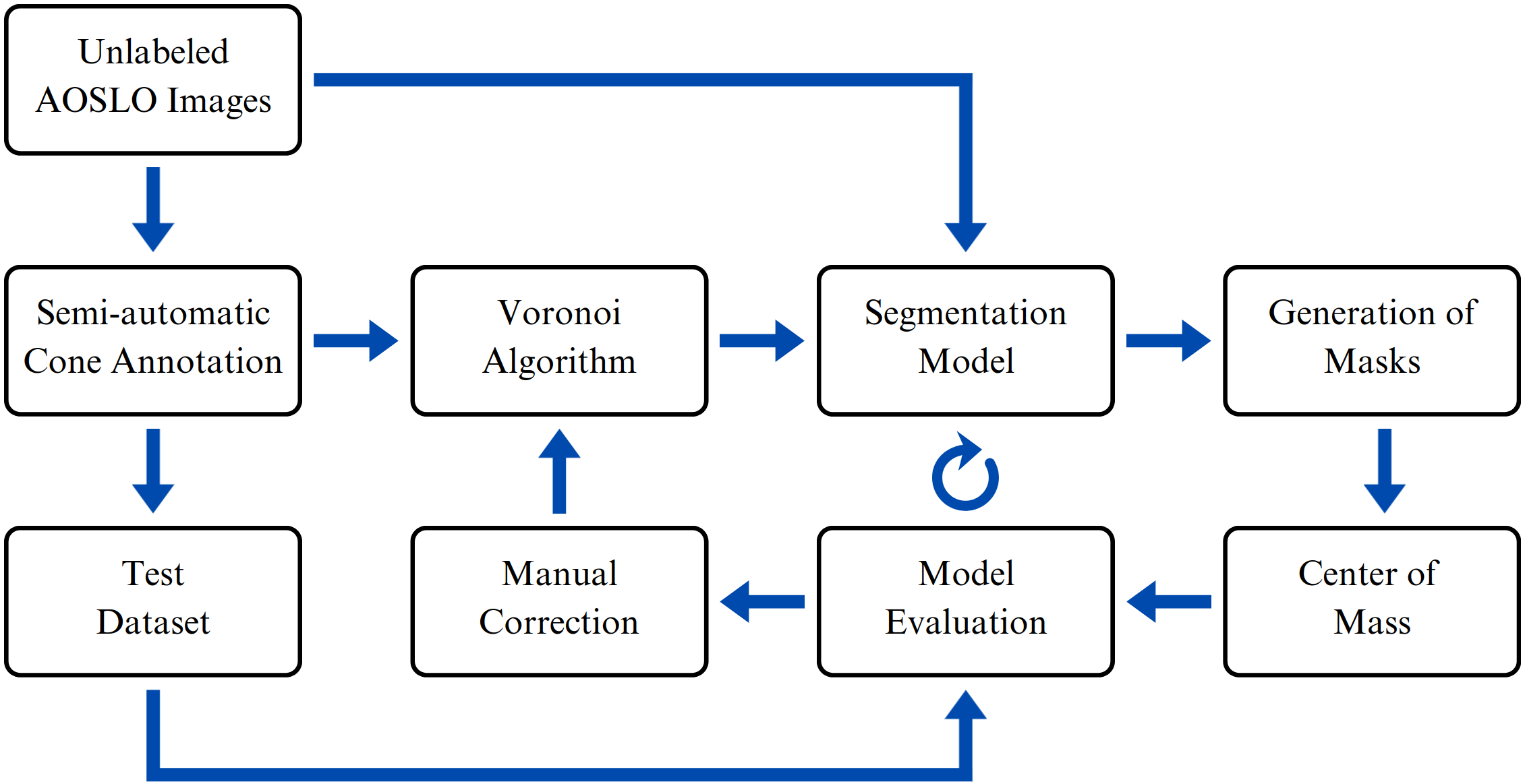}}
\end{minipage}
\caption{Pipeline of the method. Voronoi algorithm is applied to initially semi-automatically annotated cones to obtain the masks. Then, a segmentation model was trained, which generates segmentation masks for unlabeled AOSLO images. The center of mass function is applied to get the centers of the cells from segmentation masks. After each iteration step, the model was evaluated using the test subset. A manual correction step is involved in the pipeline to improve the annotations of the segmentation model of initially unlabeled images.}
\label{fig:pipeline}
\end{figure}

Fig.~\ref{fig:pipeline} illustrates the overall pipeline of the proposed method. AOSLO images were labeled and split into the test and train subsets as described in the dataset section \ref{dataset}. To all labeled areas, we applied the Voronoi algorithm to obtain the masks of the cones. Then, the human-in-the-loop step was applied: on the initially labeled AOSLO images, we trained DL-based models to generate semantic masks on unlabeled images. Then, from semantic masks, we calculated the center of mass for each segment (cone), which is basically the center of the cone we manually labeled. Therefore, we could evaluate the models by comparing the obtained centers with ground truth labels. Given the potential for initial inaccuracies, manual correction is a crucial step in the method. This step ensures the precision of the model's output by allowing the expert to review and adjust the segmented cone centers, mitigating the risk of errors in the initial automated segmentation. Adding new annotations and the manual correction step are not involved in the initial zero iteration, they are only applied starting from the first iteration step.

The Voronoi algorithm is reapplied to the refined data after correction of the centers of the cones, which was done in EXACT \cite{marzahl2021exact}. This iterative process refines the segmentation accuracy and enriches the training dataset with additional, corrected instances. Thus, the next 15\% of the total number of images is labeled at each iteration, increasing the training dataset. Each iteration concludes with an evaluation step on a test dataset to quantify the improvements. Additionally, at this stage, we apply the $K$-means algorithm for clustering the cones by the mean brightness of the center part (reflection). Therefore, we obtain the percentage of reflecting and non-reflecting cones, which is also a priori information for diagnostics. This cyclical process, encompassing both automated segmentation and expert review, ensures the development of a robust model capable of high-precision cone segmentation.

\subsection{Voronoi algorithm} 
The Voronoi algorithm is one of the more useful geometrical constructions to study point patterns since it provides all the information needed to study proximity relations between points \cite{mozos2010v}. Connecting surrounding cones and characterizing the number of sides, the Voronoi diagram allows assessment of the degree of hexagonality, and it is often used to show how disease and aging can affect this aspect of packing geometry \cite{baraas2007adaptive}. In a healthy retina, cones are packed in the most efficient manner possible, which is a hexagonal (honeycomb) arrangement. The degree of hexagonality, therefore, can be used as a proxy for general retinal health. Applying the Voronoi algorithm, we can obtain a reasonably accurate approximation of photoreceptor segmentation by labeling only the centers of cones.

\begin{figure*}[h]
  \centering
  \subfigure[]{\includegraphics[height=6cm, width=6cm]{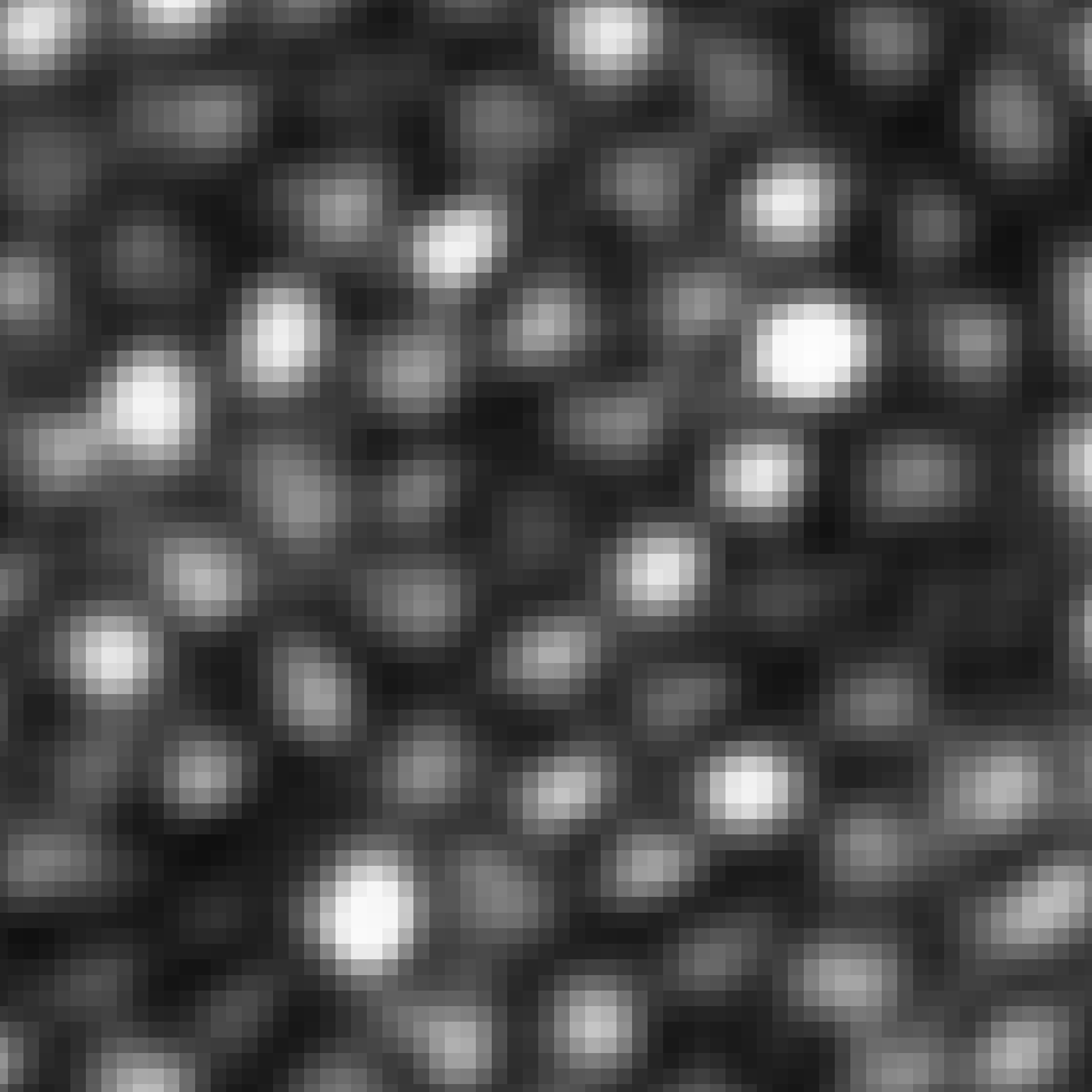}}
  \subfigure[]{\includegraphics[height=6cm, width=6cm]{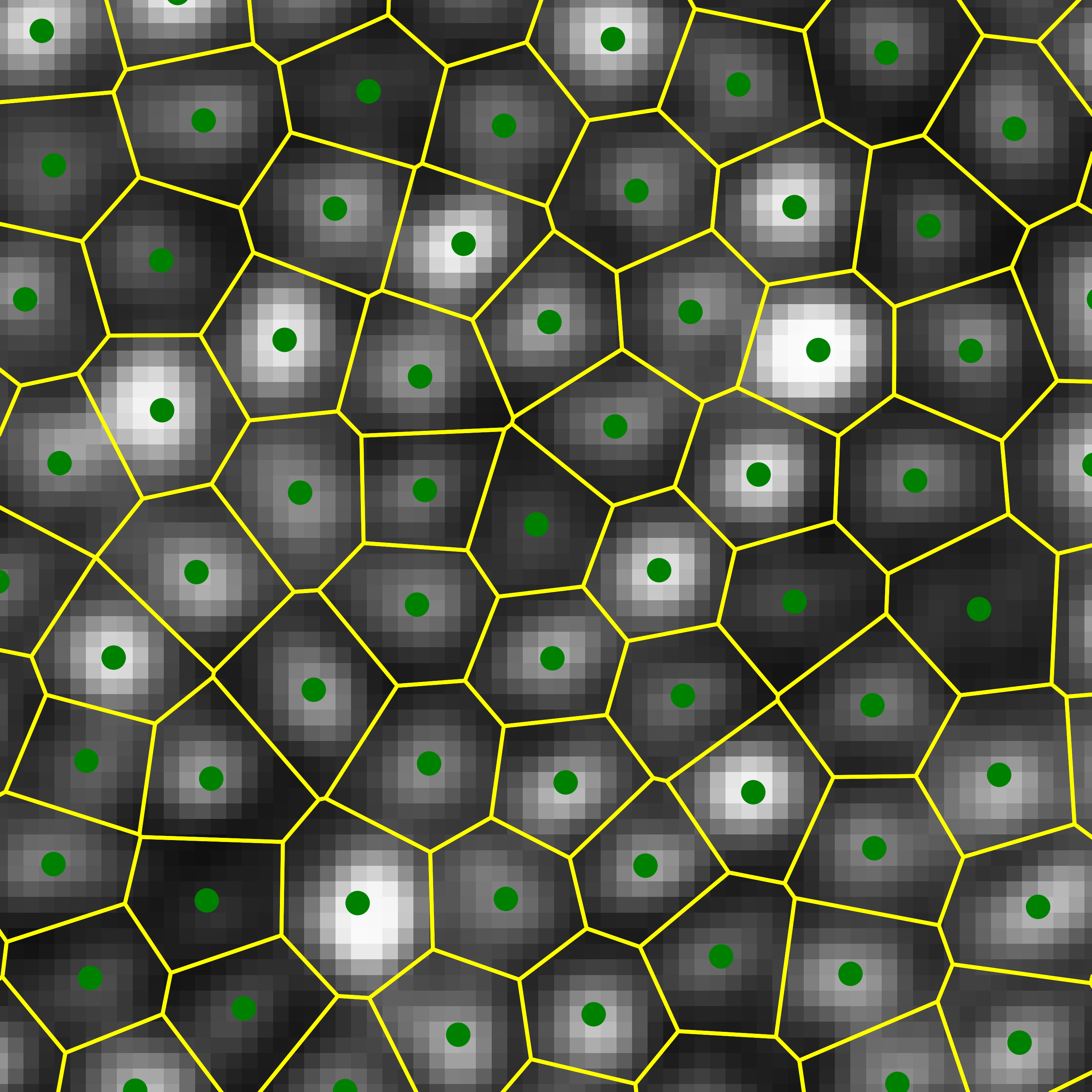}}
\caption{Application of Voronoi algorithm on the labeled AOSLO images: (a) example of the original image; (b) segmented image.}
\label{fig:iter}
\end{figure*}

\subsection{Attention-augmented U-Net} 
\begin{figure}[h]
\begin{minipage}[b]{1.0\linewidth}
  \centering
  \centerline{\includegraphics[width=12cm]{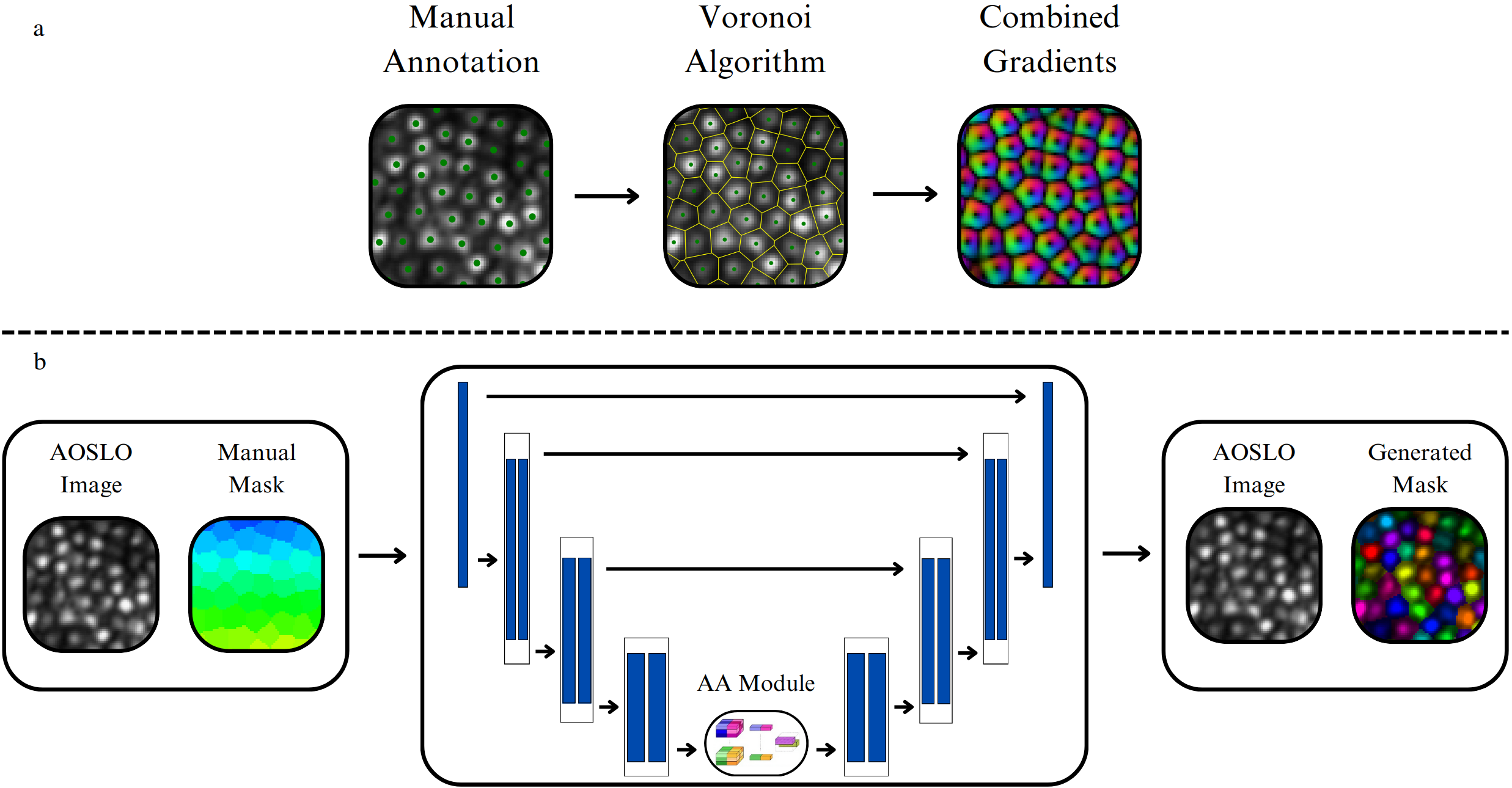}}
\end{minipage}
\caption{Model overview. (a) Transformation from the center of the cell to a gradient vector field using the Voronoi algorithm. (b) U-Net model with additional Attention-augmented module.}
\label{fig:scheme}
\end{figure}

Fig.~\ref{fig:scheme} shows the overview of the model. In our model, we applied the concept of flows (vector gradient fields) \cite{stringer2021cellpose}. This means that we trained a neural network to predict the horizontal and vertical gradients of the topological maps. Additionally, the network predicts a binary map to indicate if a given pixel is inside or outside of regions of interest. Our model was based on the general U-Net architecture \cite{ronneberger2015u} with an additional attention-augmented module (AA module) \cite{rajamani2023attention}. This module dynamically adjusts the importance of different spatial regions and channels in the input data, enabling the network to prioritize more relevant features for improved segmentation accuracy. Such augmentation facilitates precise localization and detailed segmentation in complex image datasets, which is particularly beneficial in medical imaging applications where accuracy is essential. Attention mechanisms can help the model to focus on relevant features and ignore distractions, therefore, improving segmentation accuracy. 

The AA module improves the performance of overlapping or docked objects. The nature of the Voronoi algorithm ensures the cells are always tightly packed, with no possibility of spaces in between. The AA module helps to distinguish between adjacent objects by prioritizing spatial features that define boundaries, enhancing the model's ability to separate and accurately segment individual cells.

\subsection{Center of Mass}

We calculated the center of mass to extract the centers of the cones. The center of mass is a point that corresponds to the average position of all the mass in a system. For discrete systems, the center of mass can be considered the weighted average of the positions of all elements, where the weights are the values of those elements. The cone may have several pixels corresponding to the brightest color: using the center of mass, we get the center of the brightest area.

\section{Experiments}

\subsection{Training setup.} 

The model was trained for 500 epochs on each iteration with stochastic gradient descent with a learning rate of 0.001, a momentum of 0.9, a batch size of 16 images, and a weight decay of 0.0001. All the models were trained on a single NVIDIA A100 graphics processing unit on a machine with two Intel Xeon Gold 6134 3.2 GHz and 96 GB RAM. One training iteration on this setup lasts 30 minutes, with about 10 seconds for the further inference of one image from a batch.

To predict the horizontal and vertical gradients, we used the MSE loss function. We applied the cross-entropy loss function to predict the probability that a pixel was inside or outside a cell.

\subsection{Metrics}
To match predicted points and ground truth, we applied the KDTree algorithm \cite{friedman1977algorithm}. Each predicted cell center matched with a ground truth pair was True Positive (TP), a predicted cone without a ground truth pair was False Positive (FP), and when nothing was detected where ground truth indicates a cone was a False Negative (FN) case. The $L_2$ distance ($D_{L_2}$) between pairs of points was calculated using the following formula:

\begin{equation} \label{eq:pre}
D_{L_2} = \sqrt{(x_a - x_b)^2 + (y_a - y_b)^2},
\end{equation}
where $a$ and $b$ are predicted and ground truth centers, respectively. Detected cones were evaluated using Recall, Precision, and F1-score:

\begin{equation} \label{eq:pre}
\text{Precision} = \frac{\text{TP}}{\text{TP} + \text{FP}},
\end{equation}
\begin{equation} \label{eq:rec}
\text{Recall} = \frac{\text{TP}}{\text{TP} + \text{FN}},
\end{equation}
\begin{equation} \label{eq:f1score}
\text{F}1 = \frac{2 \times \text{Precision} \times \text{Recall}}{\text{Precision} + \text{Recall}}.
\end{equation}

\subsection{Results}

Fig.~\ref{fig:iter} shows an example of the application of the algorithm on the first (a) and second (b) iterations. Green and red circles correspond to ground truth and predicted cell centers, respectively. Yellow squares on the first iteration show unlabeled cells (FNs); on the second iteration, they are correctly labeled.

\begin{figure*}[h]
  \centering
  \subfigure[]{\includegraphics[width=6cm]{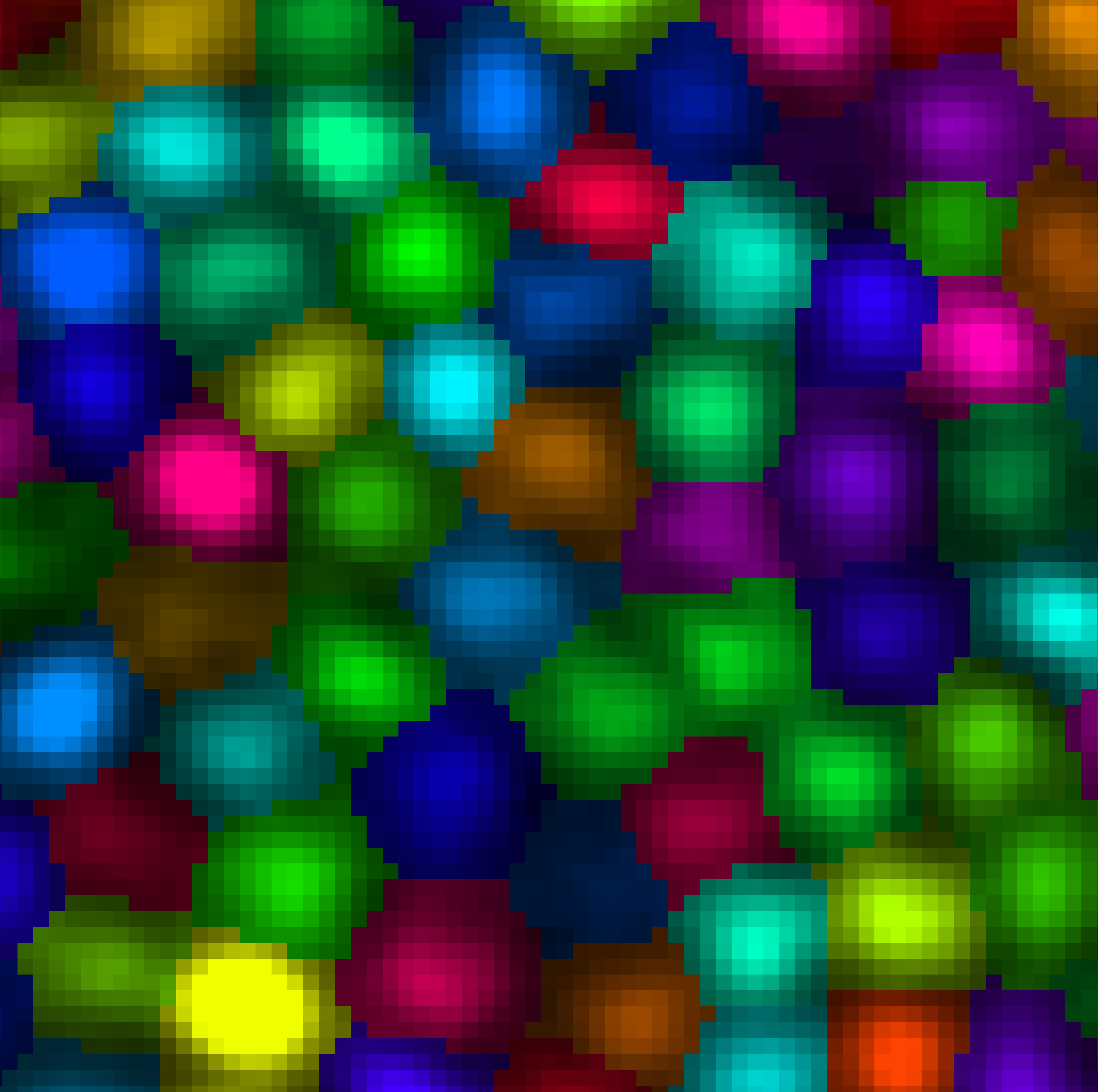}}
  \subfigure[]{\includegraphics[width=6cm]{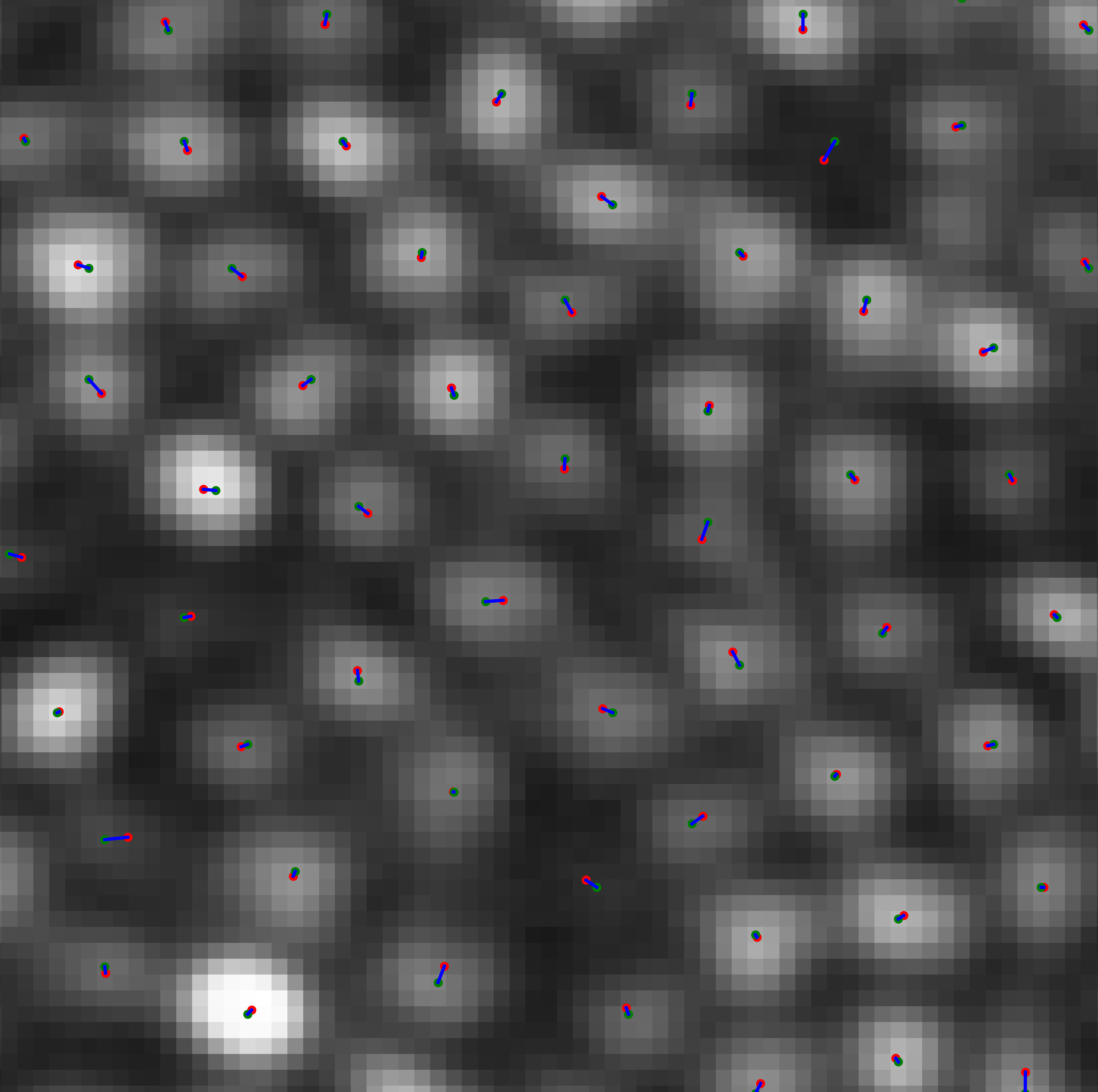}}
  \caption{Example of a predicted segmentation mask (a); example of the matching of predicted (red) and ground truth (green) centers (b). Blue lines show the $L_2$ distance.}
\label{fig:dist}
\end{figure*}

Fig.~\ref{fig:dist}a shows an example of the predicted semantic mask by our model. For this mask, the center of mass was calculated, obtaining the centers of cones that are shown in Fig.~\ref{fig:dist}b. Predicted centers (red) are matched with ground truth (green), and the distance is shown with blue connection lines.

\begin{figure*}[!htbp]
  \centering
  \subfigure[First iteration]{\includegraphics[height=8cm, width=8cm]{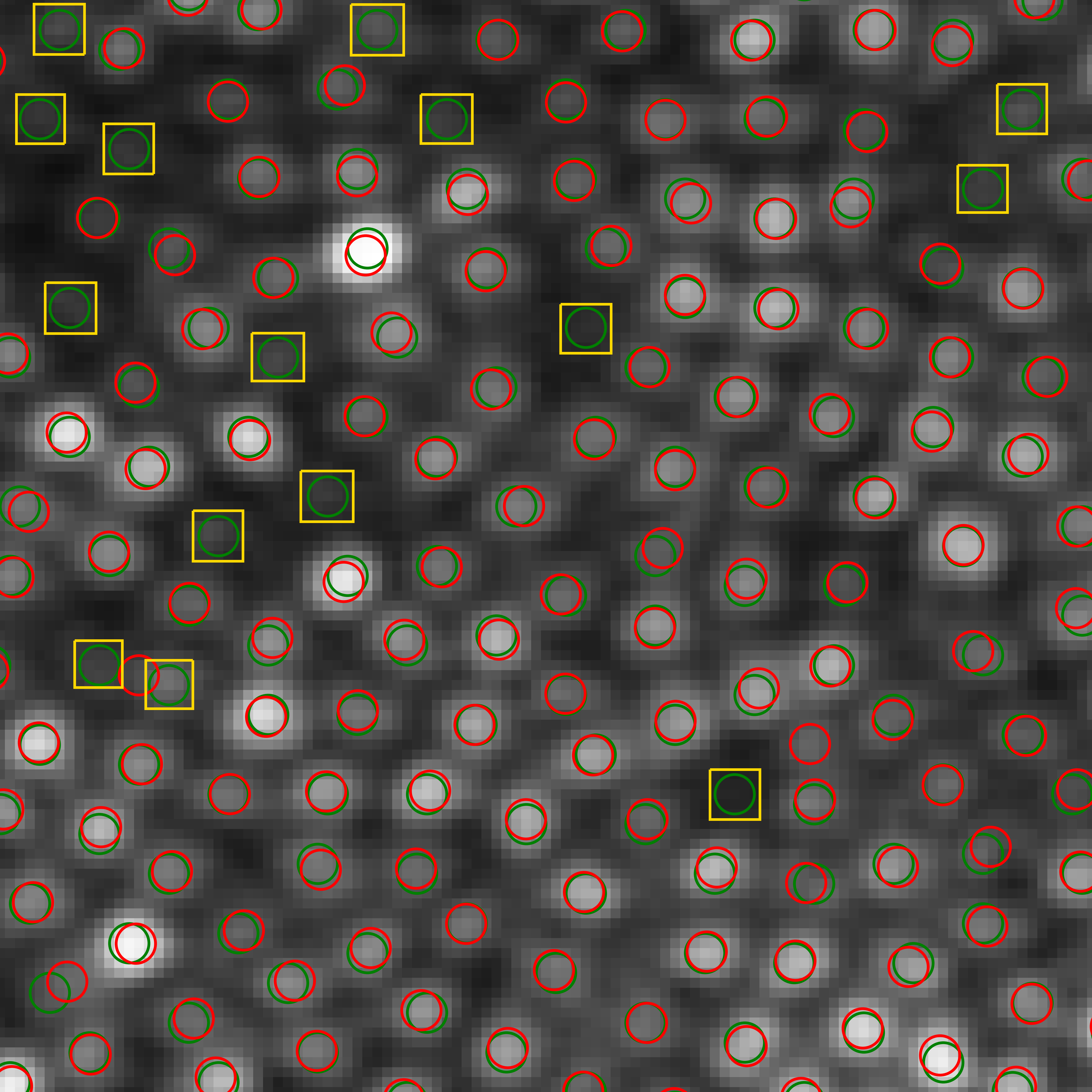}} \\
  \subfigure[Second iteration]{\includegraphics[height=8cm, width=8cm]{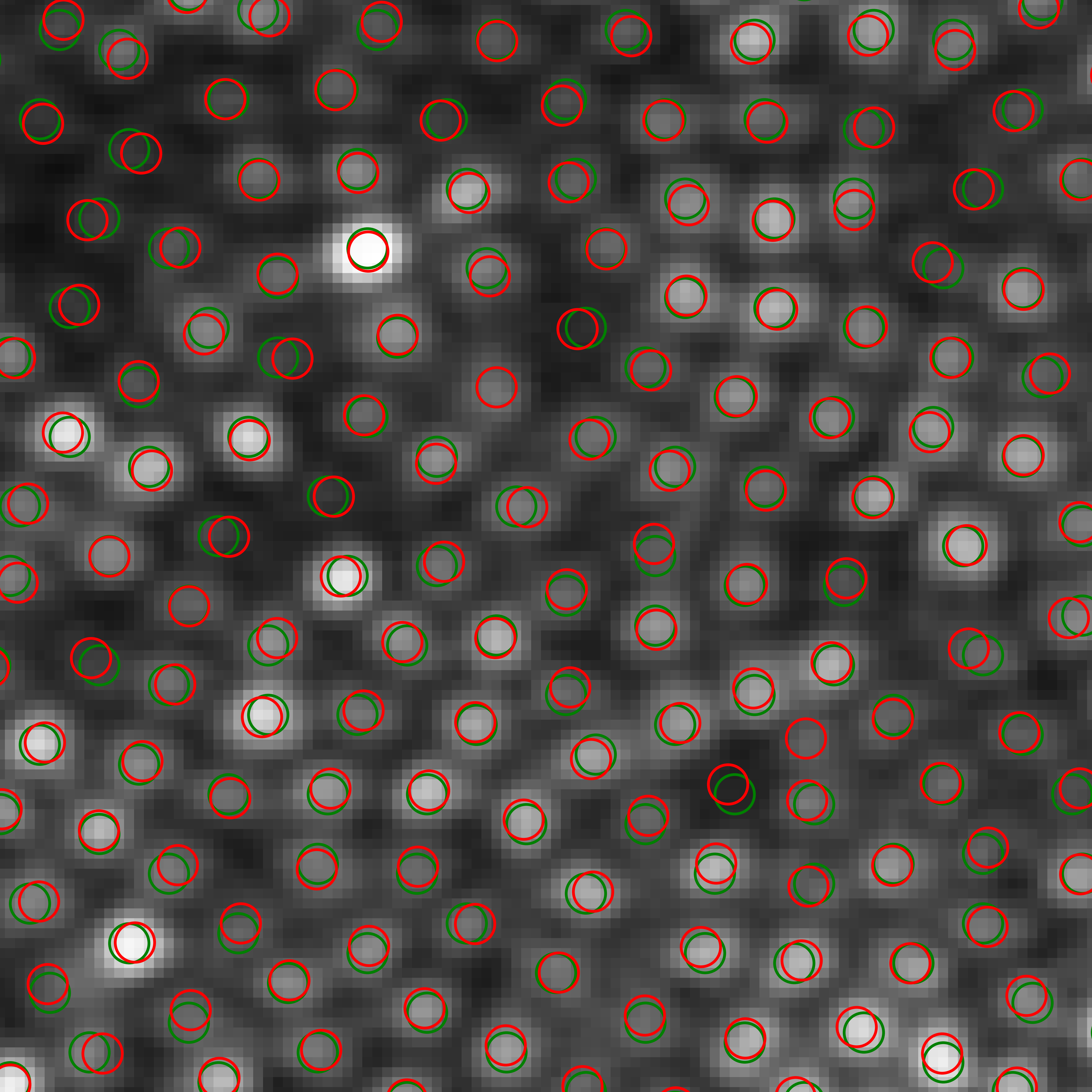}}
\caption{Evaluation of the proposed method on first (a) and second (b) iterations on a 0$^{\circ}$ test sample. Green circles correspond to the ground truth cone centers, and red to the predicted centers. Yellow squares show the False Negative predictions of the model.}
\label{fig:iter}
\end{figure*}

Table \ref{results} presents the comparative performance of the StarDist, Cellpose, and our models. Recall, Precision, and F1 score were computed separately for 0$^{\circ}$, 1$^{\circ}$, and 2$^{\circ}$ from the fovea. $D_{L_2}$ was calculated for all three degrees together. All models are characterized by a fundamental improvement after the second correction: on average, the F1 score improved by 7\%. We also see a tendency for the scores to deteriorate slightly at higher eccentricities.

\begin{table}
\setlength{\tabcolsep}{0.85 pt}
\caption{Evaluation metrics of the trained models. Best results are marked in bold.}\label{results}
\centering
\begin{tabular}{|lc|ccc|ccc|ccc|c|}
\hline
 &   &  \multicolumn{3}{c|}{0$^{\circ}$} &  \multicolumn{3}{c|}{1$^{\circ}$} &  \multicolumn{3}{c|}{2$^{\circ}$} &  All \\
\hline
Model &  It. & Recall & Precision & F1 & Recall & Precision & F1 & Recall & Precision & F1 & $D_{L_2}$ \\
\hline
StarDist & 1 & 0.833 & 0.930 & 0.879 & 0.824 & 0.919 & 0.869 & 0.810 & 0.920 & 0.861 & 7.653 \\
Cellpose & 1 & 0.843 & 0.952 & 0.895 & 0.843 & 0.941 & 0.890 & 0.811 & 0.942 & 0.872 & 7.641 \\
Ours & 1 & \textbf{0.854} & \textbf{0.953} & \textbf{0.901} & \textbf{0.846} & \textbf{0.954} & \textbf{0.897} & \textbf{0.841} & \textbf{0.955} & \textbf{0.894} & \textbf{7.637} \\
\hline
StarDist & 2 & 0.927 & 0.946 & 0.936 & 0.927 & 0.936 & 0.931 & 0.891 & 0.937 & 0.913 & 7.539 \\
Cellpose & 2 & 0.937 & 0.967 & 0.952 & 0.937 & 0.957 & 0.947 & 0.902 & 0.948 & 0.925 & 7.534 \\
Ours & 2 & \textbf{0.958} & \textbf{0.978} & \textbf{0.968} & \textbf{0.948} & \textbf{0.968} & \textbf{0.958}  & \textbf{0.940} & \textbf{0.969} & \textbf{0.954} & \textbf{7.529} \\
\hline
\end{tabular}
\end{table}

The obtained centers were clustered in terms of brightness to monitor the distribution of light-reflecting and non-reflecting (dark) photoreceptors during training. The $K$-means algorithm is a popular unsupervised machine learning technique for clustering data into a specified number of clusters, denoted by $K$. We applied the $K$-means algorithm with three clusters ($K = 3$) during each iteration for additional distribution of reflecting cone control.

\begin{figure*}[!htbp]
  \centering
  \subfigure[0$^{\circ}$]{\includegraphics[height=5cm]{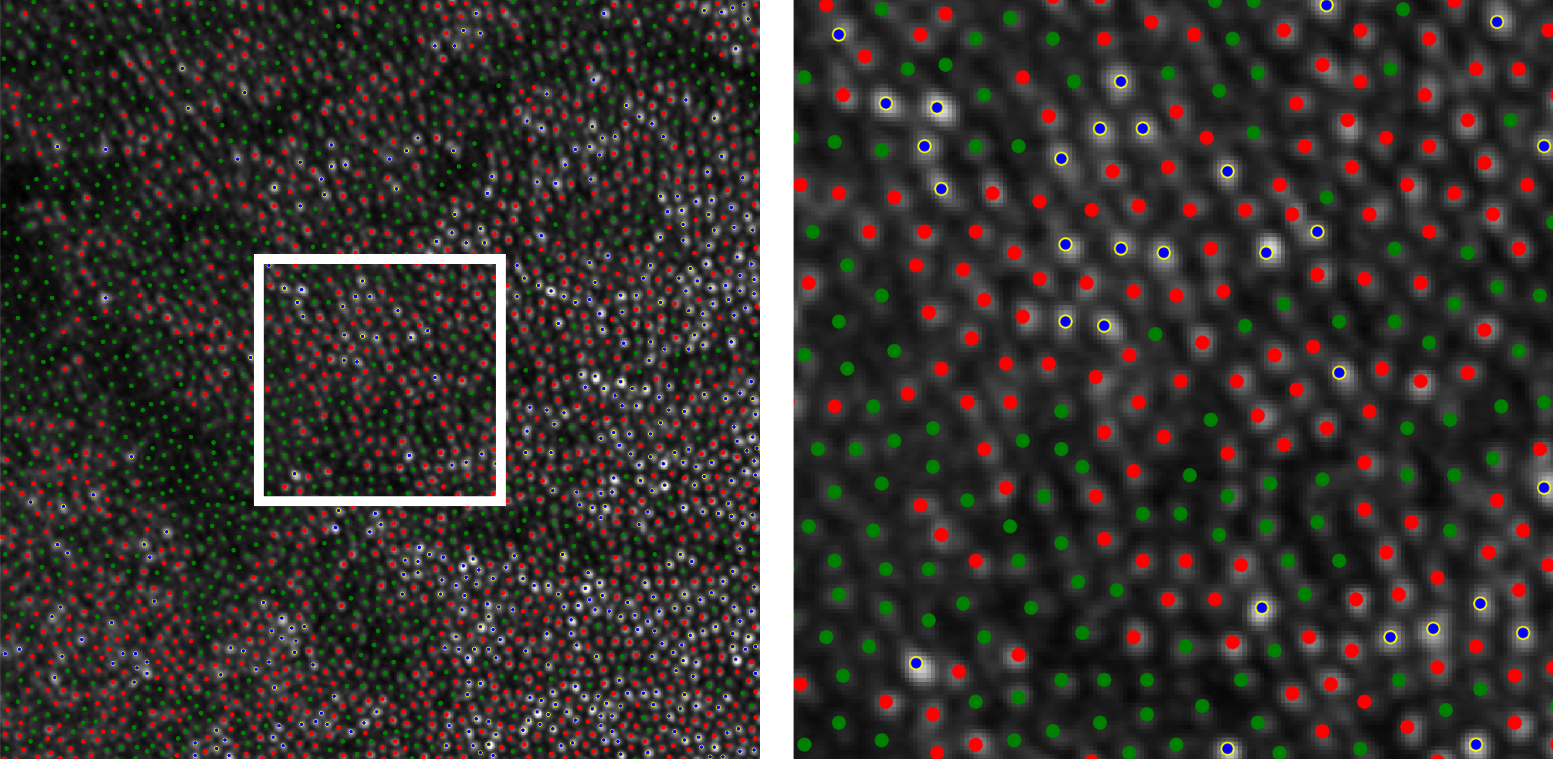}} \\
  \subfigure[1$^{\circ}$]{\includegraphics[height=5cm]{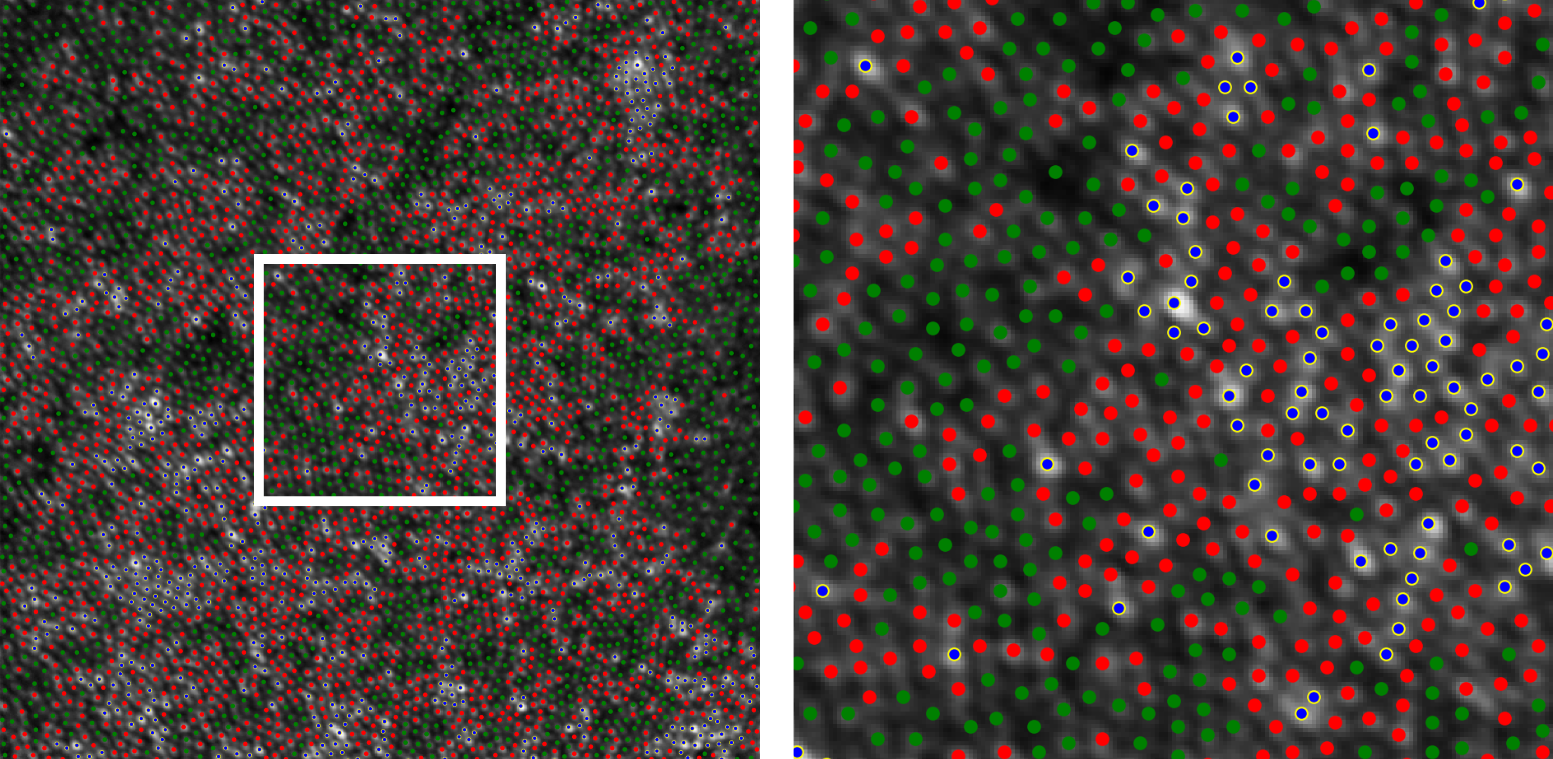}} \\
  \subfigure[2$^{\circ}$]{\includegraphics[height=5cm]{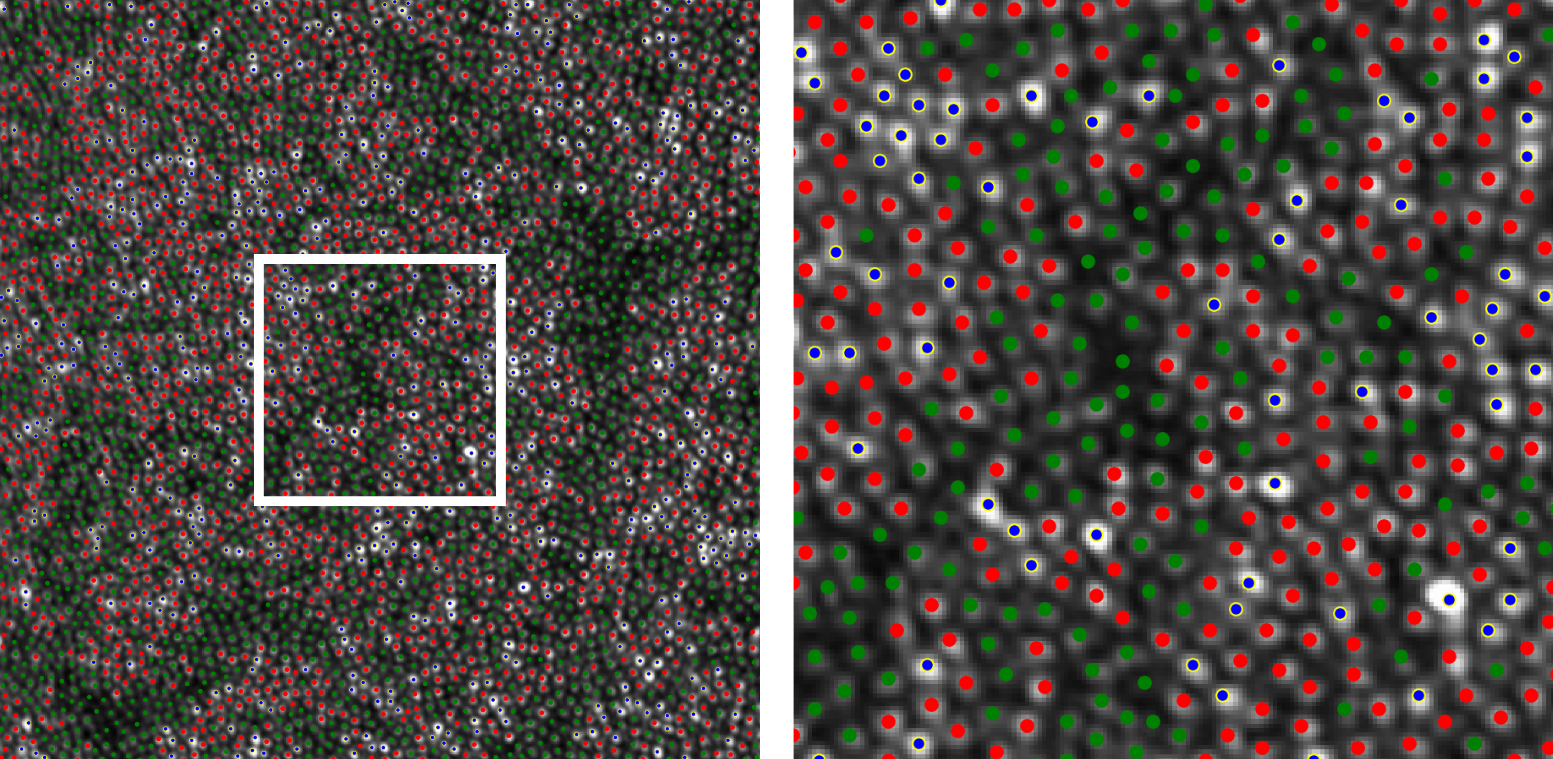}}
\caption{Examples of the method performance on the test images located at 0$^{\circ}$, 1$^{\circ}$, and 2$^{\circ}$ from the fovea with applied $K$-Means clustering algorithm. Blue-marked labels correspond to the cones with the highest reflection, green to the lowest, and red to the middle. White boxes show the location of the zoomed area in the right column.}
\label{fig:kmeans}
\end{figure*}

Clustering the cones based on their brightness level is particularly useful in retinal imaging for understanding differences between healthy and diseased retinas. Fig.~\ref{fig:kmeans} shows examples of clustering on images of 0$^{\circ}$, 1$^{\circ}$, and 2$^{\circ}$.

\begin{figure}[]
\begin{minipage}[b]{1.0\linewidth}
  \centering
  \centerline{\includegraphics[width=10cm]{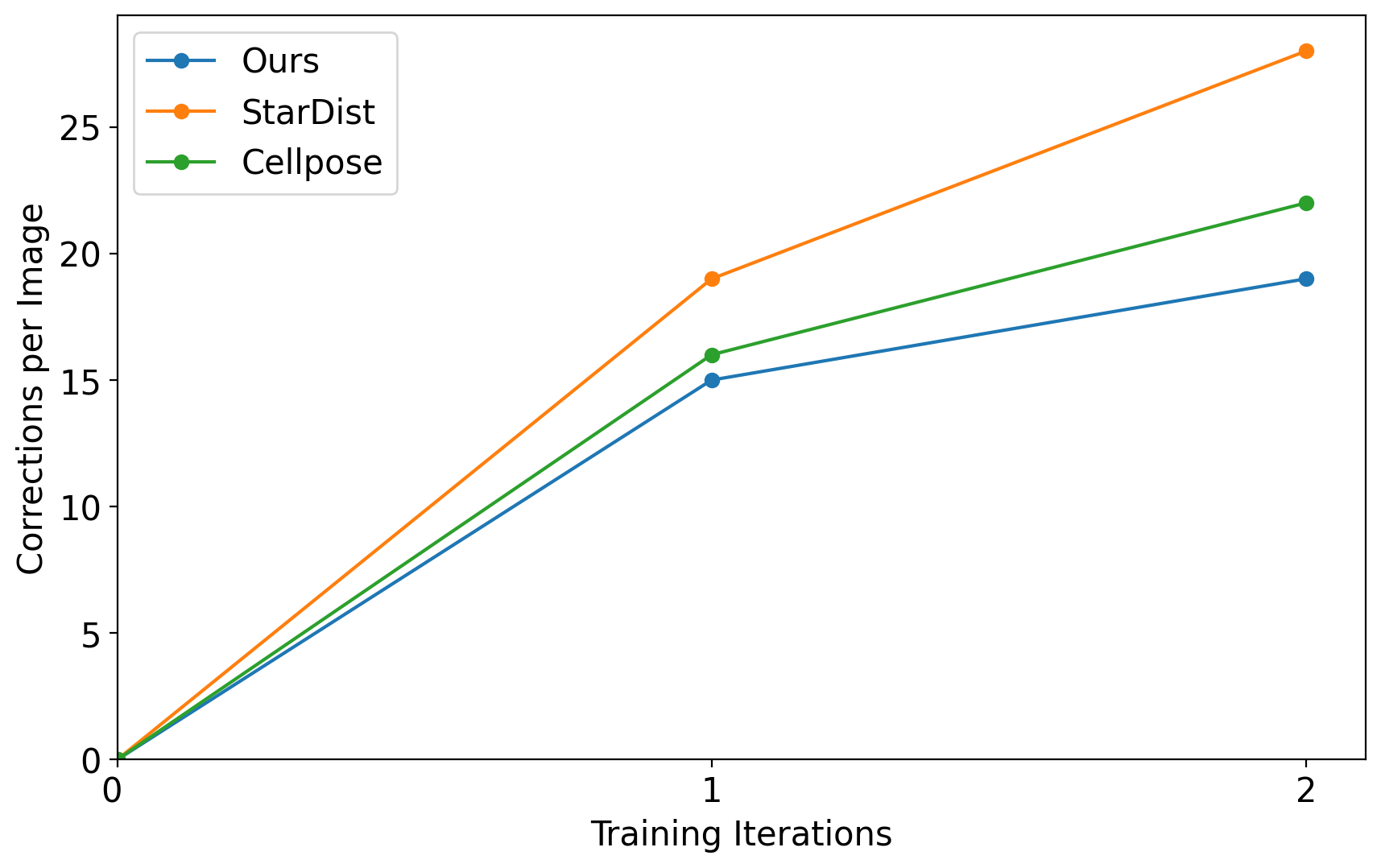}}
\end{minipage}
\caption{The cumulative average number of corrections of cone centers on the first and second iterations per image for Cellpose, StarDist, and the proposed model. The initial training iteration was done with the original labeled dataset; therefore, the number of corrections was equal to zero.}
\label{fig:iterations_graph}
\end{figure}

Fig.~\ref{fig:iterations_graph} plots the cumulative average number of corrected cone center identifications made for all three models. The values show that our proposed improvement will decrease the number of corrections for each image, compared with using Cellpose or StarDist models for the human-in-the-loop approach, potentially saving the human expert's time cost for the AOSLO cells segmentation.

\section{Conclusion}

This work describes and evaluates a method for the identification and segmentation of cone photoreceptors from AOSLO confocal images. Models were trained and tested on images covering a more extensive range of images of 18 participants with only 5\% labeled cones. Our proposed method received an overall F1 score of 0.968 for cones for 0$^{\circ}$, 0.958 for 1$^{\circ}$, and 0.954 for 2$^{\circ}$, which is better than previously reported DL approaches \cite{cunefare2018deep,cunefare2019rac}. Our method can reduce the labeling effort by requiring only a portion of labeled cones and is particularly advantageous in the ophthalmology field, where labeled data can be scarce. The work is limited to the range of degrees of eccentricity from the center from the fovea — 0$^{\circ}$, 1$^{\circ}$, and 2$^{\circ}$. Rods are already present at 1$^{\circ}$ but peaks in density at around 15$^{\circ}$; thus, rods become more and more visible on images between cones, which also require detection. Therefore, a potential improvement of the method could be to add the annotations for rods for implementing rods detection for eccentricities more than 2$^{\circ}$. This could be done using the calculated modality of the AOSLO images.

The method can be extended to the automatic identification of areas that are not cones, enabling these regions to estimate rod density. Incorporating automatic detection of inner segments in split-detection images could help to confirm that the reflected light and/or dark areas in confocal images correspond to cones. This would allow for an estimation of the number of dark cones. Furthermore, identifying retinal pigment epithelium (RPE) cells as part of this process would significantly enhance the methods' utility for clinical work and research, which leads us to future work.

\section{Code availability}
The code used to generate the results in this paper will be available at \\
\href{https://github.com/MikhailKulyabin/AOSLO}{github.com/MikhailKulyabin/AOSLO}

\section{Acknowledgments}
The authors gratefully acknowledge the scientific support and HPC resources provided by the Erlangen National High Performance Computing Center of the Friedrich-Alexander-Universität Erlangen-Nürnberg (FAU).

\bibliographystyle{splncs04}
\bibliography{refs}
%
\end{document}